\documentclass[useAMS,usenatbib,twocolumn]{mn2e}
\usepackage{graphicx,amssymb}
\usepackage{subfigure,appendix,enumerate}
\usepackage{url,hyperref,xcolor}
\usepackage{multirow}
\usepackage{lscape}
\usepackage{natbib}
\usepackage{amsmath}
\usepackage{wasysym}
\usepackage{booktabs}
\usepackage{threeparttable}
\def\apj{\ifmmode ApJ \else ApJ \fi}    % lower
\def\apjl{\ifmmode  ApJ \else ApJ \fi}    %
\def\apjs{\ifmmode  ApJS \else ApJS \fi}
\def\aap{\ifmmode A\&A \else A\&A\fi}
\def\aaps{\ifmmode A\&AS \else A\&AS\fi}    %
\def\mnras{\ifmmode MNRAS \else MNRAS \fi}    %
\def\nat{\ifmmode Nature \else Nature \fi}
\def\prl{\ifmmode Phys. Rev. Lett. \else Phys. Rev. Lett.\fi}
\def\prd{\ifmmode Phys. Rev. D. \else Phys. Rev. D.\fi}
\def\pasp{\ifmmode  PASP \else PASP \fi}
\def\pasj{\ifmmode PASJ \else PASJ \fi}
\def\gca{\ifmmode Geochimica Cosmochimica Acta \else Geochimica Cosmochimica Acta \fi}
\def\araa{\ifmmode ARAA \fi}
\def\aapr{\ifmmode A\&ARv \fi}
\def\aj{\ifmmode AJ \else AJ \fi}
\def\apss{\ifmmode Ap\&SS \else Ap\&SS \fi}

\definecolor{dark-green}{rgb}{0.1,0.49,0.4}
\hypersetup{colorlinks=true, urlcolor=blue, citecolor=dark-green}

\usepackage{color}

\begin{document}
%%%%%%%%%%%%%%%%%%%%%%%%%%%%%%%%%%%%%%%%%%%%%%%%%%%%%%%%%%%%%%%%%%%%%%
%%%%      TITLE                                                   %%%%
%%%%%%%%%%%%%%%%%%%%%%%%%%%%%%%%%%%%%%%%%%%%%%%%%%%%%%%%%%%%%%%%%%%%%%
\title[11 mHz QPO in H1743--322]{Phase-resolved spectral analysis of the 11 millihertz quasi-periodic oscillation in the black-hole candidate H1743--322} 

\author[Z. Cheng et al.]
{Zheng Cheng$^1$\thanks{Email: zheng@astro.rug.nl}, 
Mariano M\'{e}ndez$^1$, 
Diego Altamirano$^2$, 
Aru Beri$^2$,
Yanan Wang$^1$
\\ 
$^1$Kapteyn Astronomical Institute, University of Groningen, Postbus 800, 9700 AV Groningen, The Netherlands \\
$^2$School of Physics and Astronomy, University of Southampton, Southampton, Hampshire SO17 1BJ, UK \\
}

\maketitle
%%%%%%%%%%%%%%%%%%%%%%%%%%%%%%%%%%%%%%%%%%%%%%%%%%%%%%%%%%%%%%%%%%%%%%%%%%%%%
%%%%    Abstract                                                         %%%%
%%%%%%%%%%%%%%%%%%%%%%%%%%%%%%%%%%%%%%%%%%%%%%%%%%%%%%%%%%%%%%%%%%%%%%%%%%%%%
\begin{abstract}

  H1743--322 is one of the few black hole candidates (BHCs) in
  low-mass X-ray binaries that shows mHz quasi-periodic oscillations
  (QPOs) that are not associated with the more common type A, B and C
  oscillations seen in the X-ray light curves of typical BHCs systems.
  To better understand the physical origin of the mHz oscillations, we 
  carried out a phase-resolved spectroscopic study of two RXTE 
  observations of this source. As previously reported, 
  the averaged energy spectra of H1743--322 shows a strong iron line 
  at $\sim6.5$ keV. Here we found evidence that the line flux appears to be 
  modulated at twice the frequency of the mHz QPO. 
  This line flux modulation is very similar to the one previously
  found for the type-C QPO in this source. We interpret the possibly periodic line flux
  modulation with this mHz QPO in terms of Lense-Thirring precession of
  the inner flow, and discuss the possible connection with the
  modulation of the line properties with the type-C QPO frequency.

\end{abstract}
 

%%%%%%%%%%%%%%%%%%%%%%%%%%%%%%%%%%%%%%%%%%%%%%%%%%%%%%%%%%%%%%%%%%%%%%%%%%%%%
%%%%    Introduction                                                     %%%%
%%%%%%%%%%%%%%%%%%%%%%%%%%%%%%%%%%%%%%%%%%%%%%%%%%%%%%%%%%%%%%%%%%%%%%%%%%%%%
\section{Introduction}
\label{intro}

Low-mass X-ray binaries (LMXB) consist of a compact object, a neutron star or black hole, and a low-mass companion star. As the companion fills its Roche lobe, matter is transferred to the compact star through an accretion disc \citep{frank2002}. 
These systems show occasional outbursts, during which the X-ray luminosity increases by several orders of magnitude. 
These outbursts last from a few days up to months, and display distinct temporal and spectral features \citep[e.g.,][]{tanaka1995,homan2005b,remillard2006}. 

Quasi periodic oscillations (QPOs) have been found in accreting neutron-stars and black-holes systems, spanning a wide range of centroid frequencies, from milli-Hertz to $\sim1300$ Hertz \citep{klis2006book,remillard2006}. 
% Accreting neutron stars and black holes regularly display quasi periodic oscillations (QPOs) with a wide range of centroid frequencies, from milli-hertz to hundreds of Hertz \citep{klis2006book}. 
In particular, black-hole X-ray binaries (BHXRB) generally show three types of low-frequency (LF) QPOs, called types A, B, and C \citep{wijnands1999,casella2005}, and high-frequency (HF) QPOs in a few cases \citep{strohmayer2001,homan2005,belloni2012}. 
The short variability time scale of HFQPOs in BHXRBs suggests that these QPOs are produced in the innermost region of the accretion disc \citep{strohmayer2001,kluzniak2001}. 
The origin of the different types of LFQPOs is, however, not fully understood yet. 
Type-C QPOs are by far the most common type of LFQPOs among the three groups; type-C QPOs have been observed in hard states and transitions to/from the soft states with frequencies ranging from tens of mHz to $\sim$30 Hz, and always accompanied by a strong flat-top red noise component \citep{motta2015,casella2005}. 
Type-B QPOs, usually observed in the intermediate states, are strong and are accompanied by a weak red noise component \citep{belloni2005}. 
% , and are probably linked to the base of the jet \citep{fender2009}. 
There are only a handful of detections of type-A QPOs; they are detected when the source is in the soft state as weak and broad peaks, and are still poorly understood \citep{motta2016}. 
The different types of LFQPOs are believed to originate from either intrinsic luminosity variations or to geometric variability of the emission area. 
The former could, for instance, be due to fluctuations of the mass accretion rate \citep{tagger1999,cabanac2010}, shocks in the accretion flow \citep{chakrabarti1993} or intrinsic variability of the jet \citep{giannios2004}; the latter may result from Lense-Thirring precession of the accretion inner flow \citep{stella1998,ingram2009}, a precessing region related to the base of the jet \citep{fender2009,millerjones2012} or the effect of a warp in the disc \citep{fragile2008}.

% \cite{schnittman2006} found a clear correlation between the QPO properties and the binary orbit inclination that would favour the geometric explanation. 
\cite{schnittman2006} proposed a precessing ring model for the hot gas in the innermost regions of the accretion disc; in this context, the fractional rms amplitude of the type-C QPOs should be higher for high inclination system.
\cite{heil2015}, \cite{motta2015}, \cite{eijnden2017} recently reported the dependence of the QPO properties upon the source inclination using a large sample. 
For instance, \cite{motta2015} found that the type-C QPOs show higher rms amplitude for high inclination systems, while type B QPOs are stronger for low inclination ones. 
This is consistent with the idea of the Lense-Thirring precession origin of type-C QPO. 
The behaviour of type-B QPOs, on the other hand, supports the hypothesis of them being related to the jet in these sources. 

Phase-resolved spectroscopy can provide more details about the origin of LFQPOs. 
By combining the energy spectra and timing information, we can study changes of the energy spectra of a source in the QPO time scale. 
\cite{ingram2015} detected a modulation of the equivalent width of the iron line with the phase of the type-C QPO in GRS~1915+105 which, they argued, supports the geometric origin of this QPO. 
\cite{stevens2016} studied the variation of the blackbody and power-law components in the phase-resolved spectra of the type-B QPO in GX 339-4; they interpreted their findings as evidence that the type-B QPO originates at the base of the jet in this source. 
% used a special technique that fits the phase-resolved spectra in energy and time domain simultaneously on the type-B QPO in GX 339-4. 
\cite{ingram2016,ingram2017} found a modulation of the centroid energy of the iron emission line with the phase of the type-C QPO in the BHXRB H1743--322, which they regarded as strong evidence that the type-C QPO is produced by Lense-Thirring precession of the inner flow near the central compact object (see also \citealt{miller2005}). 

H1743--322 was discovered with the Ariel V All Sky Monitor by \cite{kaluzienski1977}. 
\cite{white1984} classified the source as a black-hole candidate (BHC) on the basis of its X-ray spectral characteristics. 
In March 2003, ESA's INTEGRAL satellite detected a bright and variable source, named IGR J17464-3213, in the same region in the sky \citep{revnivtsev2003}. 
Using the Rossi X-Ray Timing Explorer (\emph{RXTE}), \cite{markwardt2003} confirmed that this new transient was actually H1743--322. 
A radio counterpart was found with the Very Large Array (VLA) during the 2003 outburst by \cite{rupen2003a}, and relativistic jets were observed from this source by \cite{corbel2005} using the Australia Telescope Compact Array (ATCA). 
Due to its X-ray dipping behaviour \citep{homan2005,miller2006}, the system is believed to have a relative high inclination with respect to the line of sight ($\sim60^\circ$-$70^\circ$), and it is located at a distance of $8.5\pm0.8$ kpc \citep{steiner2012}. 

\cite{diego2012} reported the discovery of QPOs at $\sim11$ mHz in two \emph{RXTE} and one \emph{Chandra} observation of H1743--322. 
Those observations also displayed the more common type-C QPOs typically seen in this source as well as in other BHCs systems \citep{motta2015}. 
It is interesting to try and understand why this QPO is unique, and the mechanism driving the phenomenon. 
In this paper we reanalyse the \emph{RXTE} data and carry out phase resolved spectroscopy on the period of the 11 mHz QPO. 
We describe the details of the observations and the data reduction and analysis in \S\ref{data}, we show the results in \S\ref{results}, and we discuss our findings in \S\ref{discussion}. 

%%%%%%%%%%%%%%%%%%%%%%%%%%%%%%%%%%%%%%%%%%%%%%%%%%%%%%%%%%%%%%%%%%%%%%%%%%%%%
%%%%    Observations                                                     %%%%
%%%%%%%%%%%%%%%%%%%%%%%%%%%%%%%%%%%%%%%%%%%%%%%%%%%%%%%%%%%%%%%%%%%%%%%%%%%%%
\section{Data reduction}
\label{data}

The BHC H1743--322 was observed with the Proportional Counter Array (PCA, \citealt{jahoda2006}) on board the Rossi X-ray Timing Explorer (\emph{RXTE}, \citealt{bradt1993}) a total of 558 times between 2003 March and 2011 June. 
An unusual QPO at $\sim$11 mHz was detected by \cite{diego2012} twice, in the rising phase of the 2010 outburst (using \emph{RXTE} and \emph{Chandra}) and the 2011 outburst (using \emph{RXTE}). 
This QPO has only been detected in these three observations (\emph{RXTE} obsID 95368-01-01-00 and 96425-01-01-00; \emph{Chandra} ID number 401083) in this source so far. 

For our analysis we used the \emph{RXTE} PCA data of obsID 95368-01-01-00 (2010, orbits 1 and 2, GoodXenon mode) and obsID 96425- 01-01-00 (2011 orbit 3, Event mode).
We created Good Time Intervals (GTI) using the following filtering criteria: 
elevation angle larger than $10^\circ$, target offset less than 0.02$^\circ$, and two Proportional Counter Units (PCUs) on, which was the maximum number of PCUs that were active during these observations: PCU 2 and 4 were on in orbit 1 and orbit 3, and PCU 1 and 2 were on in orbit 2. 
We extracted data from all three layers of those PCUs. 
The GoodXenon mode uses two Event Analyzers (EAs) simultaneously, such that the events are split between the two EA's in two separate files. 
Therefore, we first combined the event files and then applied the GTIs to carry out the analysis. 
We extracted 1-second resolution source light curves in the 3--25 keV energy range, which we call the full-band light curves in this paper, from which we subtracted the background light curve constructed using 16 seconds resolution Standard 2 mode data after applying the dead time correction. 
As there is a gap ($\sim$3000 seconds) between the two orbits in this observation, to study the variability, we computed the Lomb-Scargle periodograms \citep{lomb1976,scargle1982,press2007} in the analysis for each of the orbits separately. 
For the second observation taken with a 64-channel Event mode, the background was subtracted using the Standard 2 mode data after rebinning them to the same number of channels as the Event mode data. 

We modelled the spectra in the 3--25 keV range using {\sc XSPEC} version of 12.9.1a. 
The interstellar absorption was considered by including the component {\sc tbabs} using the solar abundance table of \cite{wilms2000} with the photo-electric absorption cross-sections from \cite{balucinska1992} and the He cross-section by \cite{yan1998}. 
We used the component {\sc diskbb} \citep{mitsuda1984,makishima1986} to model the soft emission from the accretion disc. 
To account for the high-energy photons produced by inverse Compton scattering in a corona of hot electrons, we added a thermal Comptonisation component, {\sc nthcomp} \citep{zdziarski1996,zycki1999}. 
Compared with a power-law, the {\sc nthcomp} component gives a better description of the continuum in the bandpass where the seed photons come from; the source of seed photons in our case is the accretion disc, therefore we assumed that the seed photons have a disc-blackbody shape and linked the seed-photon temperature in {\sc nthcomp}, $kT_{\rm seed}$, to the temperature at the inner disc radius, $kT_{\rm dbb}$, in {\sc diskbb}. 
As our spectra only extend up to 25 keV, and the data showed no cut off up to that energy, we fixed the electron temperature, $kT_{\rm e}$, in {\sc nthcomp} to 1000 keV. 
The redshift of this component was fixed to 0 through out the analysis. 
We also included a Gaussian component to fit emission of an iron line that was apparent in the residuals at $\sim6.4$ keV. 

The \emph{Chandra} observation was taken in continuous-clocking (CC) mode, in which the background spectrum cannot be accurately extracted;
the properties of the line, on the other hand, are very sensitive to the shape of the continuum above $\sim7-8$ keV, where the background emission is comparable to the source. 
Because of this, and the impossibility of obtaining an accurate background spectrum, we did not use the \emph{Chandra} data further, as the line parameters cannot be retrieved from this observation. 
To check whether the \emph{Chandra} data were generally consistent with the \emph{RXTE} observation, we subtracted a rough model of the background to the data and fitted the spectrum with the same model we used for the \emph{RXTE} spectra. 
Unlike in the case of the \emph{RXTE} spectra, the parameters of the Gaussian component cannot be well constrained by the \emph{Chandra} data. 
We therefore fixed the centroid energy and the width of the Gaussian component to the value we found in the \emph{RXTE} observations, and fitted only the normalisation of the line. 
We confirmed that the 95\% upper limit of the flux of the line in the \emph{Chandra} data is consistent with the value we measured in the \emph{RXTE} observations.

%%%%%%%%%%%%%%%%%%%%%%%%%%%%%%%%%%%%%%%%%%%%%%%%%%%%%%%%%%%%%%%%%%%%%%%%%%%%%
%%%%      Analysis & Results                                             %%%%
%%%%%%%%%%%%%%%%%%%%%%%%%%%%%%%%%%%%%%%%%%%%%%%%%%%%%%%%%%%%%%%%%%%%%%%%%%%%%
\section{Results}
\label{results}

\subsection{Average Spectra}

\begin{table*}
\begin{tabular}{l c c c}
\hline
Model Parameters & Orbit 1 & Orbit 2 & Orbit 3\\
\hline
$N_{\rm H}$ (cm$^{-2}$)& \multicolumn{3}{c}{1.8$\times10^{22}$ (fixed)}\\
$kT_{\rm dbb}$ (keV) & 1.35$\pm$0.07 & 1.35$\pm$0.06 & 1.35$\pm$0.07\\
$N_{\rm dbb}$ & 4.90$_{-0.48}^{+0.67}$ & 4.95$_{-0.39}^{+0.53}$ & 3.91$_{-0.41}^{+0.54}$\\
$\Gamma$ & 1.60$\pm$0.01 & 1.59$\pm$0.01 & 1.53$\pm$0.01 \\
$kT_{\rm e}$ (keV) & \multicolumn{3}{c}{1000.0 (fixed)}\\
$N_{\rm nth}$ & 0.13$\pm$0.01 & 0.13$\pm$0.01 & 0.11$\pm$0.01 \\
$E_{\rm line}$ (keV) & 6.53$\pm$0.09 & 6.39$\pm$0.05 & 6.49$\pm$0.08 \\
$\sigma$ (keV) & 0.48$\pm$0.16 & 0.39$\pm$0.14 & 0.39$\pm$0.13 \\
$N_{\rm line}$ ($\times10^{-3}$) & 1.32$\pm$0.31 & 1.35$\pm$0.25 & 1.19$\pm$0.22 \\
\hline
$F_{\rm X}$ ($10^{-9}$ erg cm$^{-2}$ s$^{-1}$) & 2.80$\pm$0.01 & 2.87$\pm$0.01 & 2.64$\pm$0.01\\ 
$F_{\rm dbb}$ ($10^{-9}$ erg cm$^{-2}$ s$^{-1}$) & 0.14$\pm$0.02 & 0.14$\pm$0.02 & 0.11$\pm$0.02\\
$F_{\rm nth}$ ($10^{-9}$ erg cm$^{-2}$ s$^{-1}$) & 2.64$\pm$0.02 & 2.72$\pm$0.02 & 2.51$\pm$0.02\\
$F_{\rm line}$ ($10^{-11}$ erg cm$^{-2}$ s$^{-1}$) & 1.38$\pm$0.32 & 1.39$\pm$0.25 & 1.24$\pm$0.23\\
\hline
$\chi^2$/d.o.f. & \multicolumn{3}{c}{1.35/113}\\
\hline
\end{tabular}
\caption{Fitting results to the average spectra of H1743--322 for the three orbits simultaneously. 
The seed-photon temperatures in {\sc nthcomp}, $kT_{\rm seed}$, were linked to the temperatures at the inner disc radius, $kT_{\rm dbb}$. 
All errors correspond to 1-$\sigma$ confidence levels. 
% The hydrogen column density, N$_{\rm H}$ and electron temperature of {\sc nthcomp}, kT$_{\rm e}$ were fixed to 1.8$\times10^{22}$ cm$^{-2}$ and 1000 keV, respectively.
}
\label{t1}
\end{table*}

We initially fitted the average spectra for the three orbits simultaneously using only the thermal Comptonisation model plus a soft disc component. 
Because the PCA instrument is not sensitive below 3 keV, the hydrogen column density could not be well constrained using our data. 
Different values of $N_{\rm H}$ have been previously reported for this source: $N_{\rm H}= 1.35 \times10^{22} \rm{cm}^{-2}$ \citep{ingram2016}, $N_{\rm H}= 2.01 \times10^{22} \rm{cm}^{-2}$ \citep{stiele2016} and $N_{\rm H}= 2.2 \times10^{22} \rm{cm}^{-2}$ \citep{mcclintock2009}. 
\cite{prat2009} used 1.8 $\times10^{22} \rm{cm}^{-2}$, which they determined from \emph{Swift} and \emph{XMM-Newton} observations. 
\cite{chaty2015} agreed with this value, as they suggested that a relative low absorption would give spectral indices more compatible with thermal emission in the soft state. 
We therefore fixed $N_{\rm H}=1.8 \times10^{22} \rm{cm}^{-2}$, but since our analysis focused on the spectra above 3 keV, especially the energy range of the iron emission line, our results are not affected by the choice of the value of $N_{\rm H}$. 

We could not get an acceptable fit using the model {\sc tbabs*(diskbb+nthcomp)} : the best fit yielded a $\chi^{2}$ of 418.8 for 121 degrees of freedom; the residuals indicated the need of an extra component at around 6.5 keV, consistent with emission from the iron K-$\alpha$ line, possibly coming from reflection off the accretion disc. 
We therefore added a Gaussian component to our model, allowing the centroid energy to vary between 6 and 7 keV; on the other hand, the width and the normalisation were left free and allowed to vary in each orbit.
The fit with the Gaussian line yielded a $\chi^{2}$ of 152.4 for 113 degrees of freedom. 
An F-test yields a null-hypothesis probability of $1.4\times10^{-21}$, indicating that the Gaussian component is significantly required by the fit. The best-fitting parameters are shown in Table~\ref{t1}.

\subsection{Phase resolved spectra}

\begin{figure}
\begin{center}
\includegraphics[width=7.9cm]{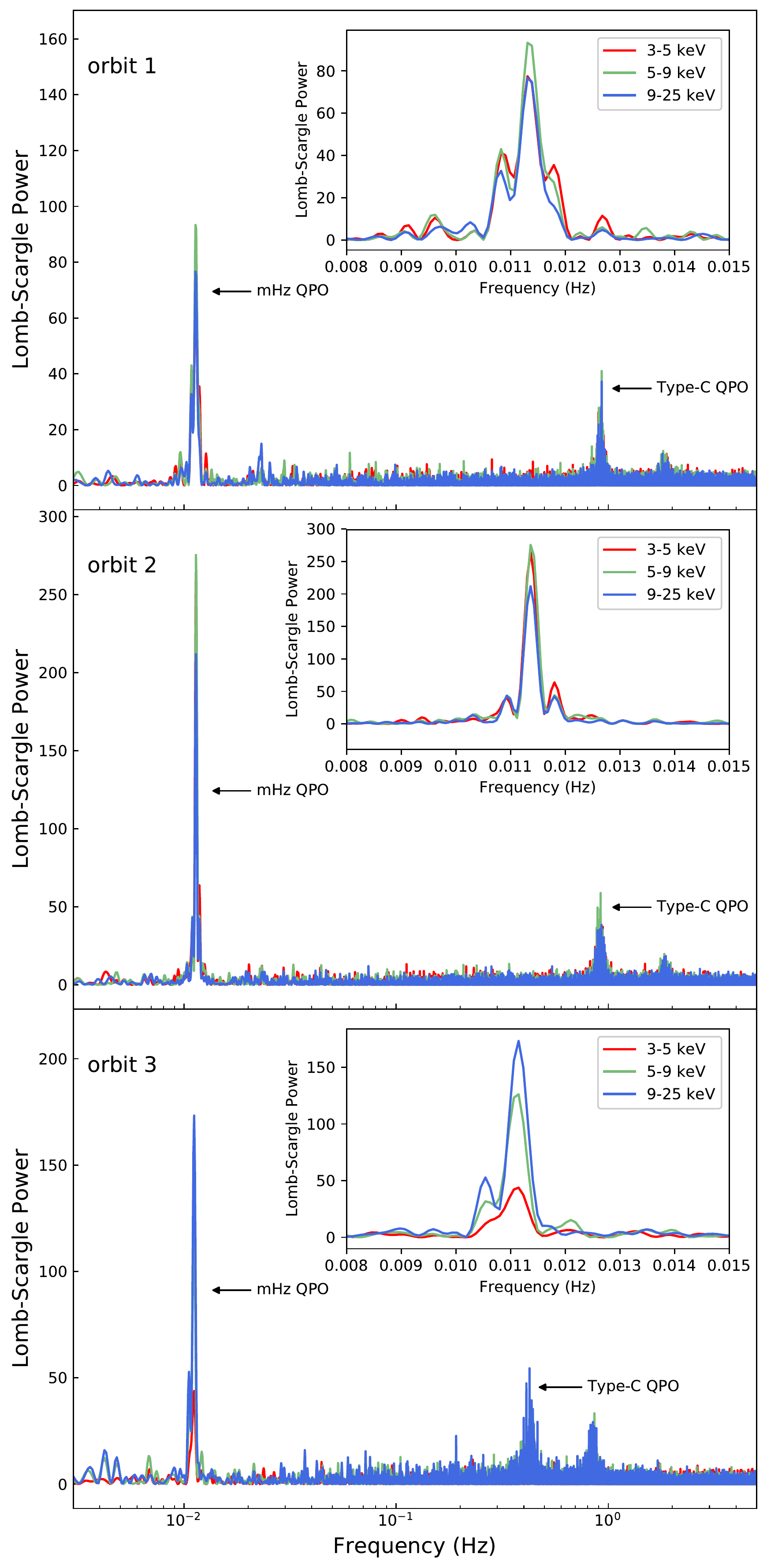}
\caption{
Lomb-Scargle periodograms of the three orbits of H1743--322 for three energy bands. 
The inset plot in each panel shows a zoom of the frequency axis from 8 to 15 mHz. 
The red, green and blue lines indicate the Lomb-Scargle periodograms in the 3-5 keV, 5-9 keV and 9--25 keV bands, respectively. 
The centroid frequency of the full-band mHz QPOs are 11.3 mHz, 11.4 mHz and 11.1 mHz for the three orbits, respectively. 
The type-C QPO and its second harmonic are also apparent in this figure. 
The fundamental frequency of the full-band type-C QPO in the three orbits is 0.92 Hz, 0.92 Hz and 0.43 Hz, respectively. 
Neither the frequency of the mHz QPOs or the type-C QPOs show any energy dependence. 
}
\label{p1}
\end{center}
\end{figure}
We computed the Lomb-Scargle periodogram of the light curves of the three orbits in three different energy bands, 3--5 keV, 5--9 keV and 9--25 keV; 
we found that, in each orbit, the frequency of the mHz QPO is consistent with being the same in all three energy bands. 
We show the Lomb-Scargle periodogram of the three orbits in Figure~\ref{p1}. 
Both the $\sim11$ mHz QPOs and the type-C QPOs (with its harmonic) can be seen in the figure. 
The false-alarm probability for mHz QPOs in these three orbits, taking into account the number of trials, is $1.4\times10^{-24}$, $1.8\times10^{-81}$, and $2.9\times10^{-21}$, respectively. 
The frequency of the mHz QPOs is quite stable; the centroid frequency of the mHz QPOs for the three orbits measured in the full-band periodogram are at $11.3\pm0.2$, $11.4\pm0.2$ and $11.1\pm0.2$ mHz, respectively. 
The frequency of the type-C QPO, however, is a factor of $\sim2$ lower in orbit 3 compared with the first two orbits. 
The frequency of the type-C QPO for the three orbits are $0.92\pm0.03$ Hz, $0.92\pm0.04$ Hz and $0.43\pm0.02$ Hz, respectively.  

\begin{figure}
\begin{center}
\includegraphics[width=6cm,angle=270]{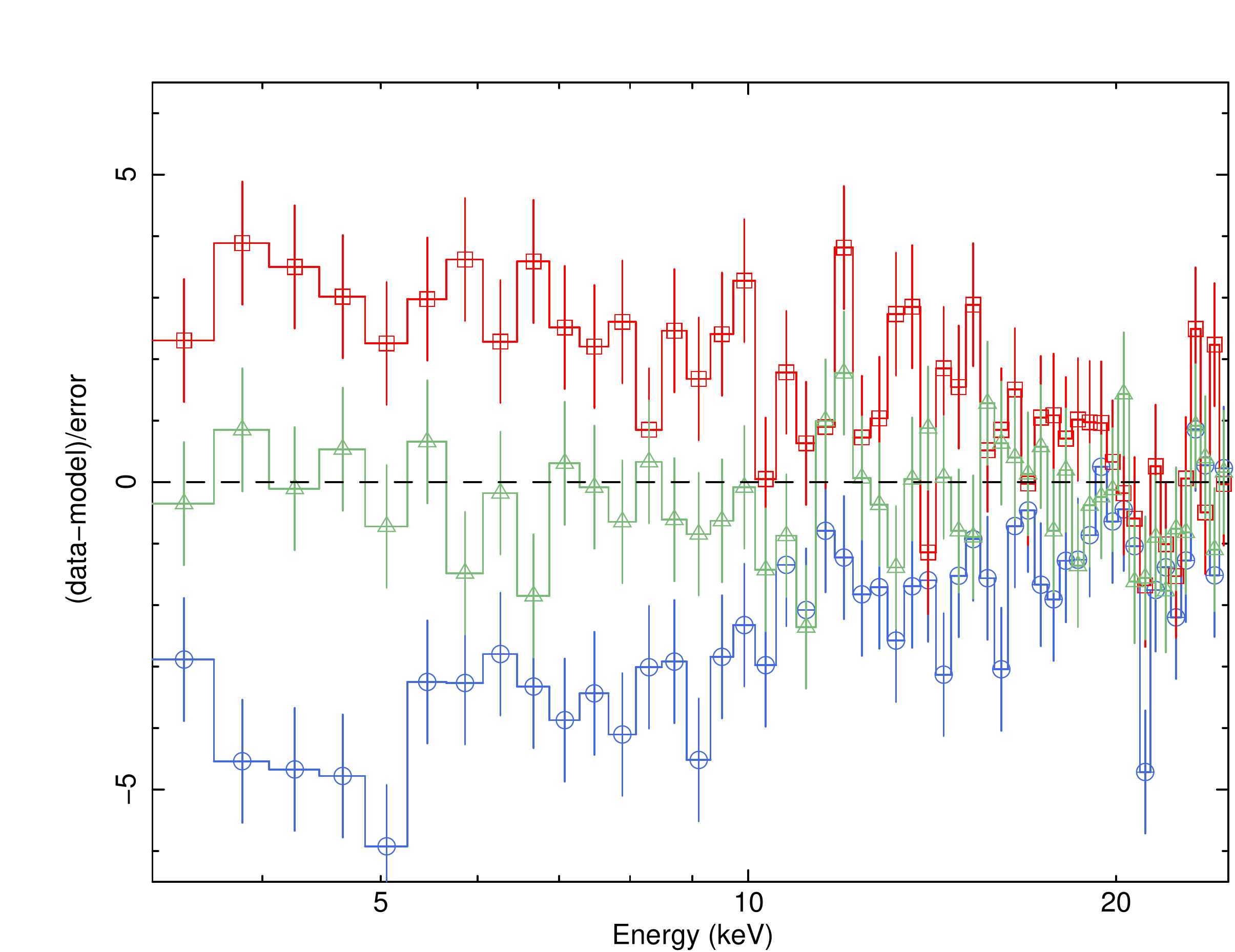}
\caption{
Deviations from the average spectra of H1743--322 in orbit 1 for three different QPO phase intervals. The blue circles, red squares and green triangles indicate the deviations from the average spectrum at, respectively, $\phi = 0.0$--$0.1$, corresponding to the minimum of the light curve folded at the QPO frequency, $\phi=0.5$--$0.6$, corresponding to the maximum, and $\phi=0.9$--$1.0$, corresponding to the mean value of the light curve.
}
\label{p2}
\end{center}
\end{figure}

Based on the QPO frequency, we folded the background-subtracted light curves and created the average oscillation waveforms for the three orbits separately. 
As the observation is in the rising phase of the outburst, there is a significant increase of the count rate between the two orbits in the first observation. 
Therefore, instead of combining the spectra of these two orbits together, we analysed the spectra for each of the orbits separately.

We divided the QPO cycle into ten phase bins with the same duration for each orbit, 
where we defined the phase $\phi=0.0-0.1$ at the first minimum of the folded light curve (see below). 
We then computed the GTIs for each of the phase bins and extracted the phase resolved spectra using the {\sc ftool} command {\sc seextrct}. 

For each orbit separately, we used the model {\sc tbabs*(diskbb+nthcomp+gaussian)} to fit the ten phase-resolved spectra simultaneously. 
We first tried to let the Gaussian line centroid energy and normalisation free to change between phases, however, the 1-$\sigma$ errors of the best-fitting energy of the line were $\sim0.2-0.5$ keV, which are too large to detect the expected variability, c.f. $\sim0.3$ keV as found by \cite{ingram2016}, 
given also that the energy resolution of the PCA at 6 keV is $\sim1$ keV.  
% is not good enough to constrain the energy variation of the line between phases, 
For the rest of the analysis we decided to link the centroid energy and width of the line to be the same in all phases and let the normalisation free to change.

In Figure~\ref{p2} we plot the spectra of three selected QPO phases in the first orbit, phase 0.0--0.1, 0.5--0.6 and 0.9--1.0 of the QPO cycle, respectively, from which we subtracted the average spectra. 
From this Figure it is apparent that the spectra show more variability in the soft than in the hard part. 
The fitting results also support this idea: the contribution of the soft component ({\sc diskbb}) changes from 2\% of the total flux between 3--25 keV in phase 0.0--0.1 to 9\% of the total flux in phase 0.6--0.7, 
% The changes of parameters in the soft component are more prominent between phases, 
whereas the photon index of the {\sc nthcomp} component in the different phases is roughly consistent with the value obtained from fitting the average spectra. 
The coupling of the {\sc nthcomp} and {\sc diskbb} components cause the parameters to be less well constrained, resulting in large errors of the fitting parameters. 
Because of this, we decided to link the photon index and the normalisation of {\sc nthcomp} component across phases. 

\begin{figure*}
\begin{center}
\includegraphics[width=16cm]{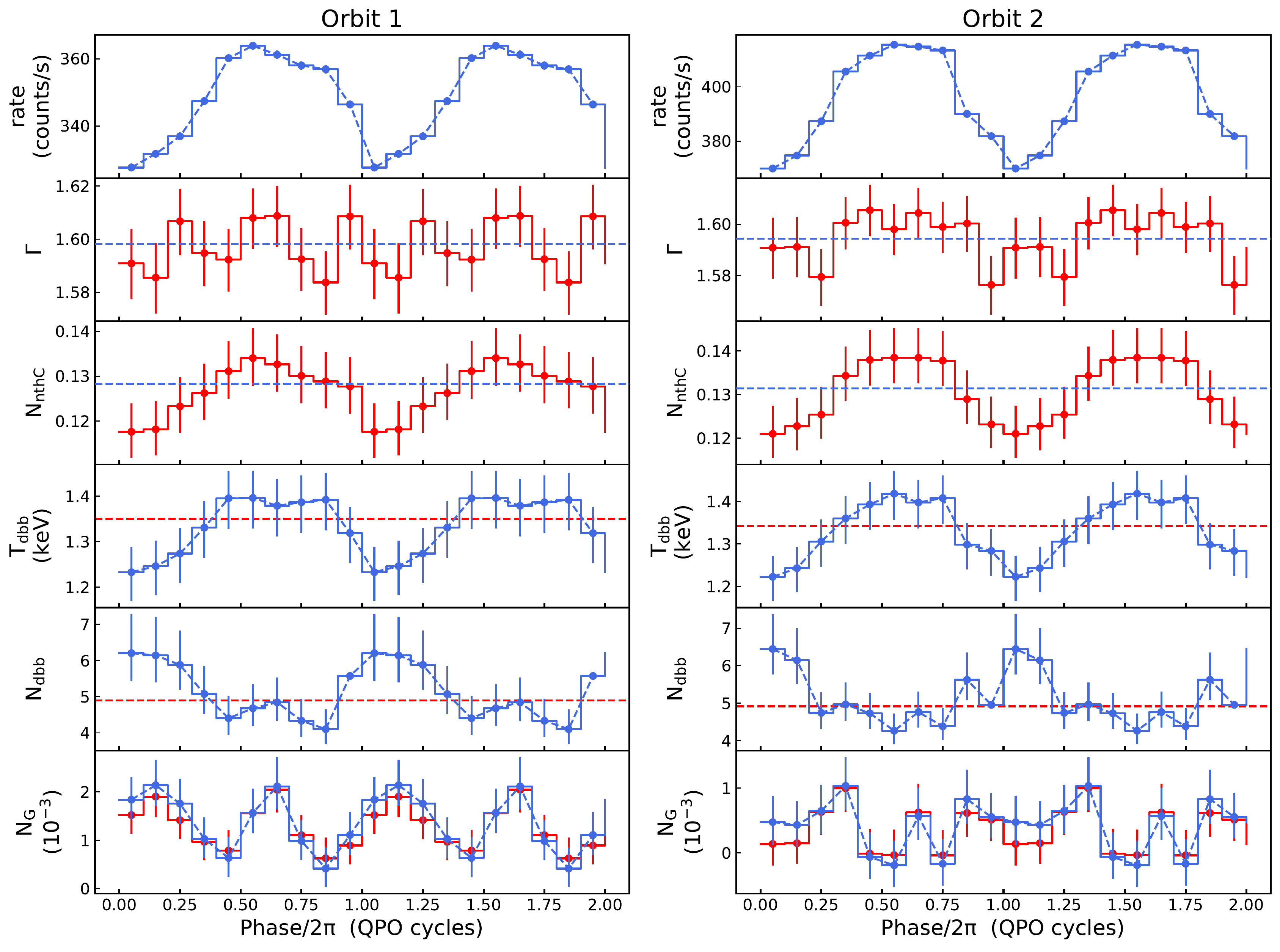}
\caption{Best-fitting parameters of the mHz QPO phase-resolved spectra of H1743--322. 
The abscissa is the phase of the mHz QPO in units of 2$\pi$; for clarity  we plot two cycles. 
The blue symbols in the lower three panels correspond to the fitting results when the parameters of the hard component were linked across phases. 
The red symbols in the second, third and bottom panels correspond to the fitting results when the parameters of the soft component were linked across phases. 
The left panels show the fitting result for orbit 1. 
The parameters from top to bottom are: the net count rate, the photon index and the normalisation of {\sc nthcomp}, the temperature at the inner radius of the disk, $kT_{\rm dbb}$, the normalisation of the disk blackbody, and the normalisation of the Gaussian line. 
The right panels show the same fitting parameters for orbit 2. 
All errors represent the 1-$\sigma$ confidence interval of the given parameter. 
}
\label{p3}
\end{center}
\end{figure*}

In Figure~\ref{p3}, we show the fitting parameters as a function of the QPO phase for the two orbits of observation 95368-01-01-00. 
The change of the disc temperature follows the modulation of the count rate in both orbits, whereas the {\sc diskbb} normalisation changes in the opposite way. 
A clear modulation of the Gaussian normalisation is also apparent in the first orbit; 
this modulation has a period of half the period of the QPO, with peaks at phase 0.1--0.2 and 0.6--0.7 of the QPO cycle. 
The fits yield a $\chi^{2}$ of 526.0 for 486 degrees of freedom, with the best-fitting centroid energy and width of the line being 6.55$\pm$0.07 keV, 0.49$\pm$0.15 keV, respectively. 
If we fit the phase-resolved spectra by linking all the Gaussian normalisations to the same value, the best fit yields a $\chi^{2}$ 552.4 for 495 degrees of freedom. 
The F-test yields a significance of 2.9$\sigma$ for this variability. 

Even though the source shows more variability in the soft part of the spectrum, the soft component contributes less than $\sim5$\% of the total flux in the time averaged spectrum, and the hard component dominates the spectrum during the full QPO cycle. 
For this reason, we also fitted the data linking the temperature and normalisation of the {\sc diskbb} component between different phases, letting the parameters of the {\sc nthcomp} component free. 
In this case the fits yield a $\chi^{2}$ of 523.2 for 486 degrees of freedom, and the best-fitting centroid energy and width of the line are 6.56$\pm$0.07 keV, 0.45$\pm$0.15 keV, respectively. 
The photon index is consistent with the value we got from the time averaged spectra within errors, and the normalisation of the Gaussian line shows exactly the same behaviour as in the case when we linked the parameters of the hard component. 
The F-test yields a significance of 1.9$\sigma$ for the variability of the line in this case. 
The results of these fits are also shown in Figure~\ref{p3}. 

We get consistent results, but with larger errors bars, if we let both the photon index and normalisation of {\sc nthcomp} free; the best fit yields $\chi^{2}=$ 501.9 for 468 degrees of freedom for orbit 1, and $\chi^{2}=$ 480.2 for 498 degrees of freedom for orbit 2. 
In this case an F-test indicates that the significance level of variability of the Gaussian normalisation in orbit 1 is 1.6$\sigma$. 
The Gaussian normalisation does not show a similar modulation in the second and third orbit.

\begin{figure}
\begin{center}
\subfigure{\includegraphics[width=8cm]{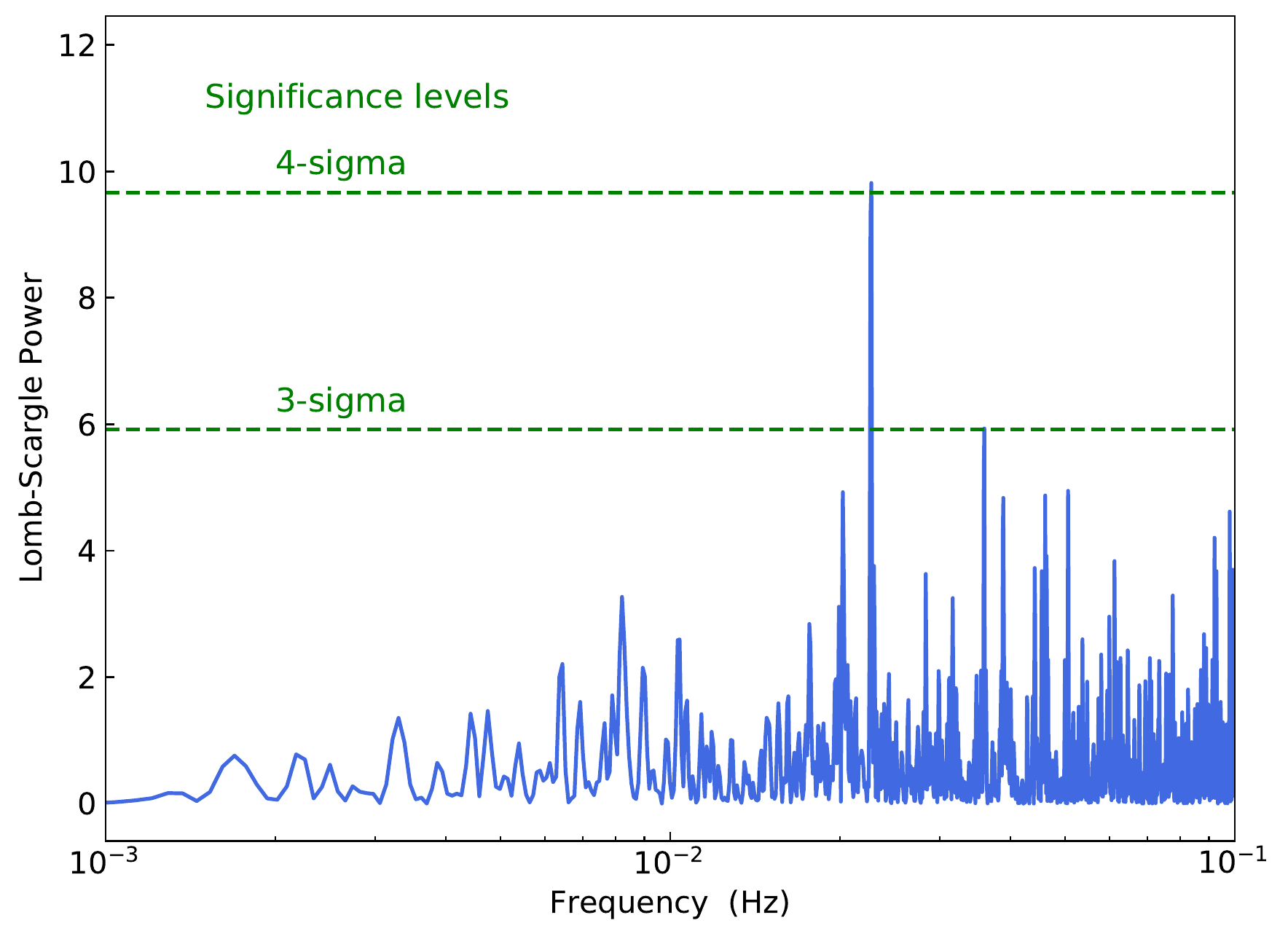}}
\caption{
Lomb-Scargle periodogram of the iron emission line flux of H1743--322 in orbit 1. 
The light curve in the energy range of the iron line, 5.5--7.5 keV, was subtracted from the light curve just above the energy range of the line, 7.5--12.0 keV light curve. 
The two horizontal green dashed lines indicate the 3$\sigma$ and 4$\sigma$ single-trial False-alarm probability levels, respectively. 
The peak of Lomb-Scargle power is at 0.0227 Hz, which is twice the frequency of the mHz QPO in orbit 1.}
\label{p6}
\end{center}
\end{figure}

In order to confirm that the modulation is not affected by the choice of the model, we first excluded the energy range between 5--8 keV of the phase-resolved spectra, and fitted the spectra with a model without the iron emission line, {\sc tbabs*(diskbb+nthcomp)}. 
We got the best-fitting parameters by letting the temperature and normalisation of {\sc diskbb}, and the photon index and normalisation of {\sc nthcomp} free to change across phases. 
We then fitted the full energy range (3--25 keV) spectra with a model that included an extra Gaussian component. 
All the parameters of the continuum components were fixed at the best-fitting values we found when we excluded the 5--8 keV range. 
The centroid energy and the width of the Gaussian component were linked to be the same, while the normalisation of the Gaussian component was free to change, as a function of phase. 
The Gaussian normalisation shows the same modulation, and an F-test yields a significance of 3.9$\sigma$ for this variability. 
Additionally, to confirm such modulation in a model-independent way, we extracted two light curves, one between 5.5--7.5 keV (the mean count rate is 86.5$\pm$0.2 counts/s) to include the energy range of the line, and the second one in the 7.5--12.0 keV energy band (the mean count rate is 84.9$\pm$0.2 counts/s). 
We subtracted one light curve from the other to remove the contribution of the continuum, and then computed the Lomb-Scargle periodogram of the subtracted light curve; the result is shown in Figure \ref{p6}. 
There is a peak in the Lomb-Scargle periodogram at 0.0227 Hz, which is at twice the frequency of the 11.3 mHz QPO shown in Figure \ref{p1}. 
This is consistent with the modulation of the Gaussian normalisation found in the first orbit. 
The single trial false-alarm probability of this modulation is 5$\times$$10^{-5}$. 
This shows that the flux modulation of the iron emission line is model-independent and highly significant. 

\begin{figure}
\begin{center}
\includegraphics[width=8cm]{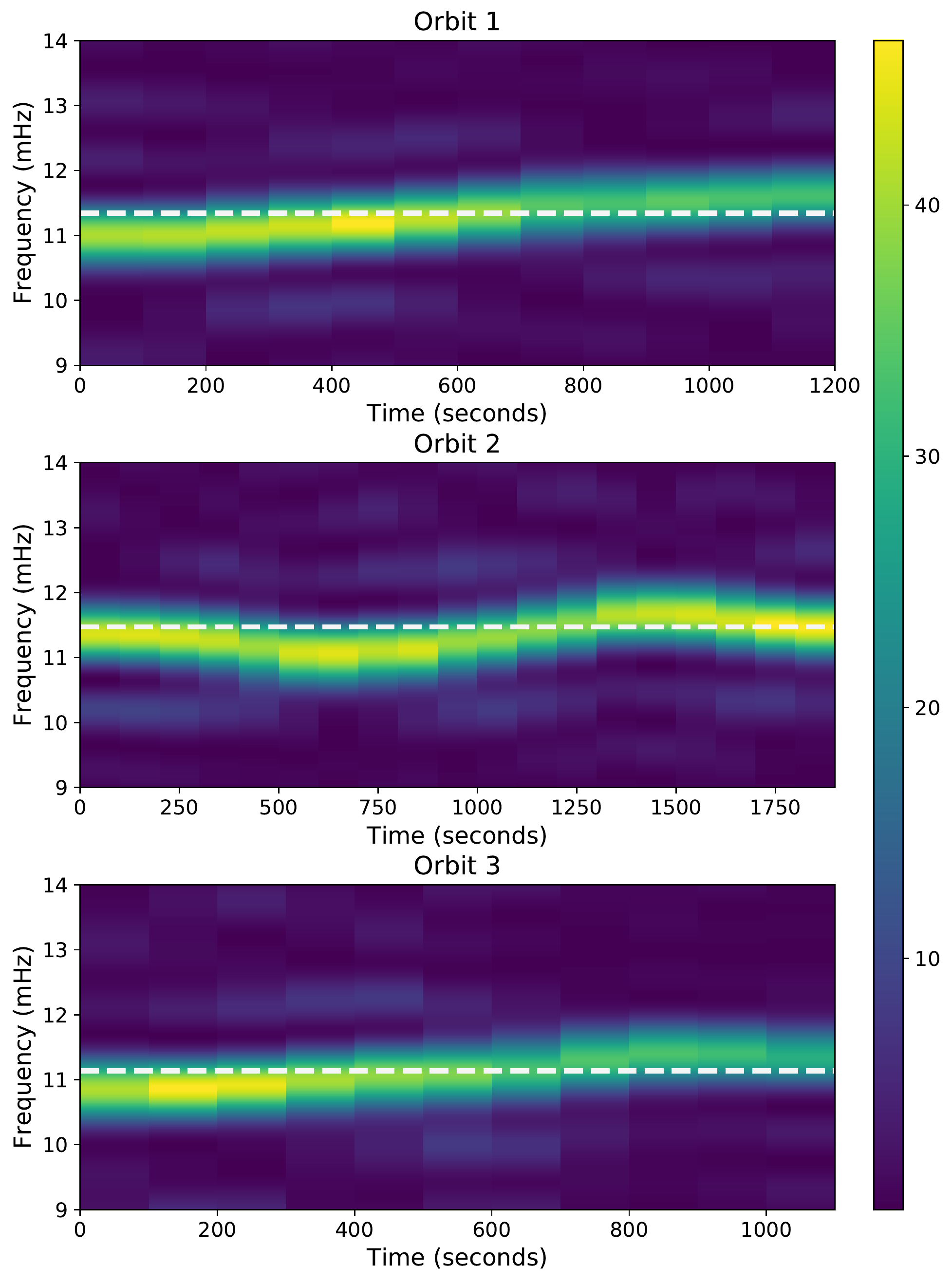}
\caption{Dynamic power spectra of H1743--322 for the three \emph{RXTE} orbits. The colour indicates the Lomb-Scargle power at the corresponding frequency in a sliding `data window' of 1200 seconds width. 
The horizontal axis marks the start time of the `data window' which was shifted 100 seconds each time. 
The white dashed line indicates the frequency of the peak Lomb-Scargle power for the full dataset of each orbit.}
\label{p4}
\end{center}
\end{figure}

To explore whether the modulation of the Gaussian normalisation is present only during parts of orbits 2 and 3, and whether the modulation changed during orbit 1, 
% Due to the absence of the modulation of Gaussian normalisation in the second orbit, 
we computed a dynamical power spectrum (DPS) of the three orbits. 
For this we calculated the Lomb-Scargle periodograms in a sliding window of 1200 seconds, shifted by 100 seconds each time to produce periodograms for a series of overlapping time intervals for the three orbits separately.  
The DPS are shown in Figure~\ref{p4}. 
The colour indicates the Lomb-Scargle power at the corresponding frequency, while the white dashed line indicates the frequency that we used to create the phase-resolved spectra in Figure~\ref{p3}. 
The QPO frequency in all orbits is relatively stable; variations are less than 10\%, yet in the second orbit the QPO frequency shows slightly more variability compared to the first and the third orbits. 

\begin{figure*}
\begin{center}
\subfigure{\includegraphics[width=8cm]{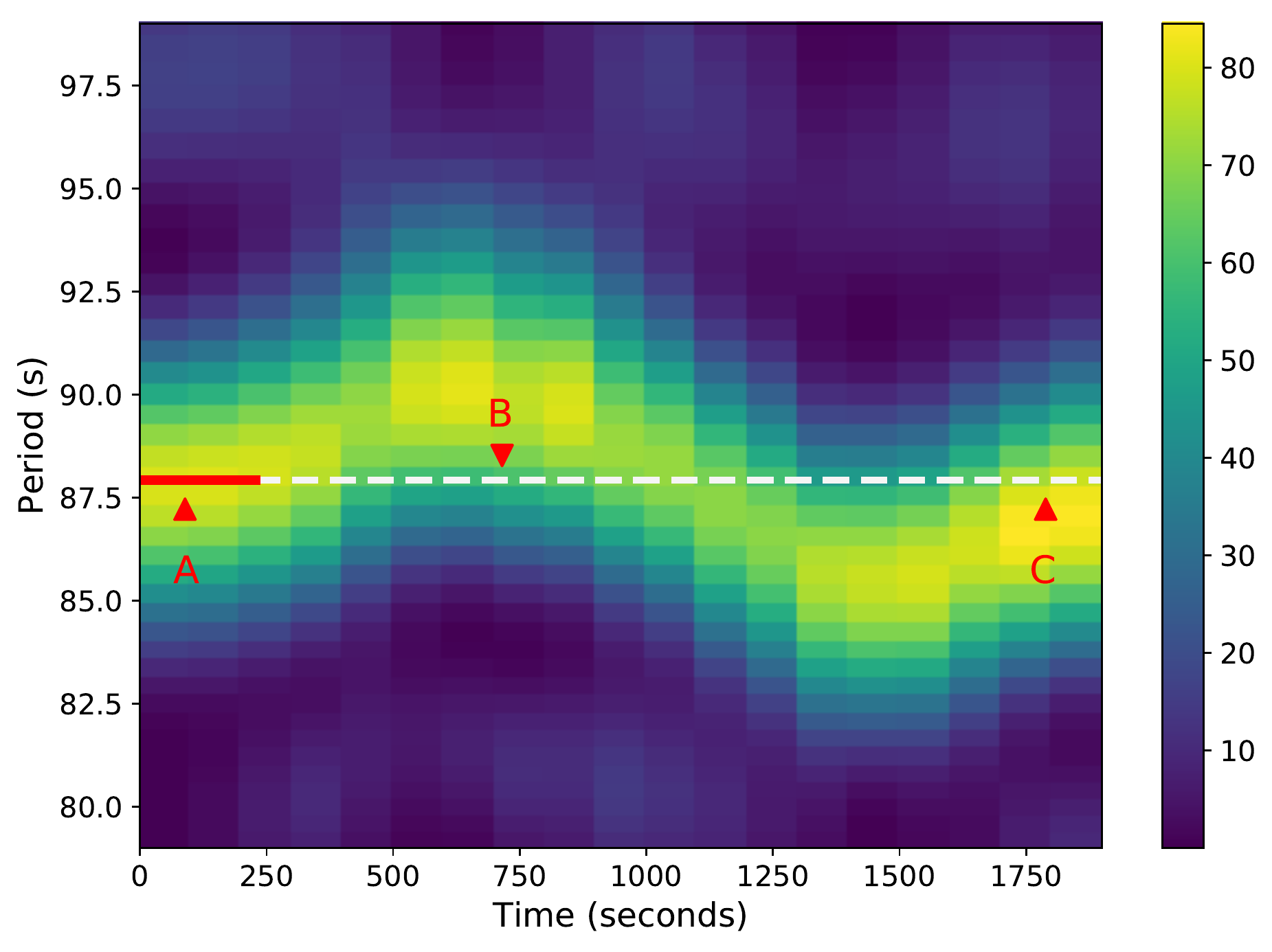}}
\subfigure{\includegraphics[width=8cm]{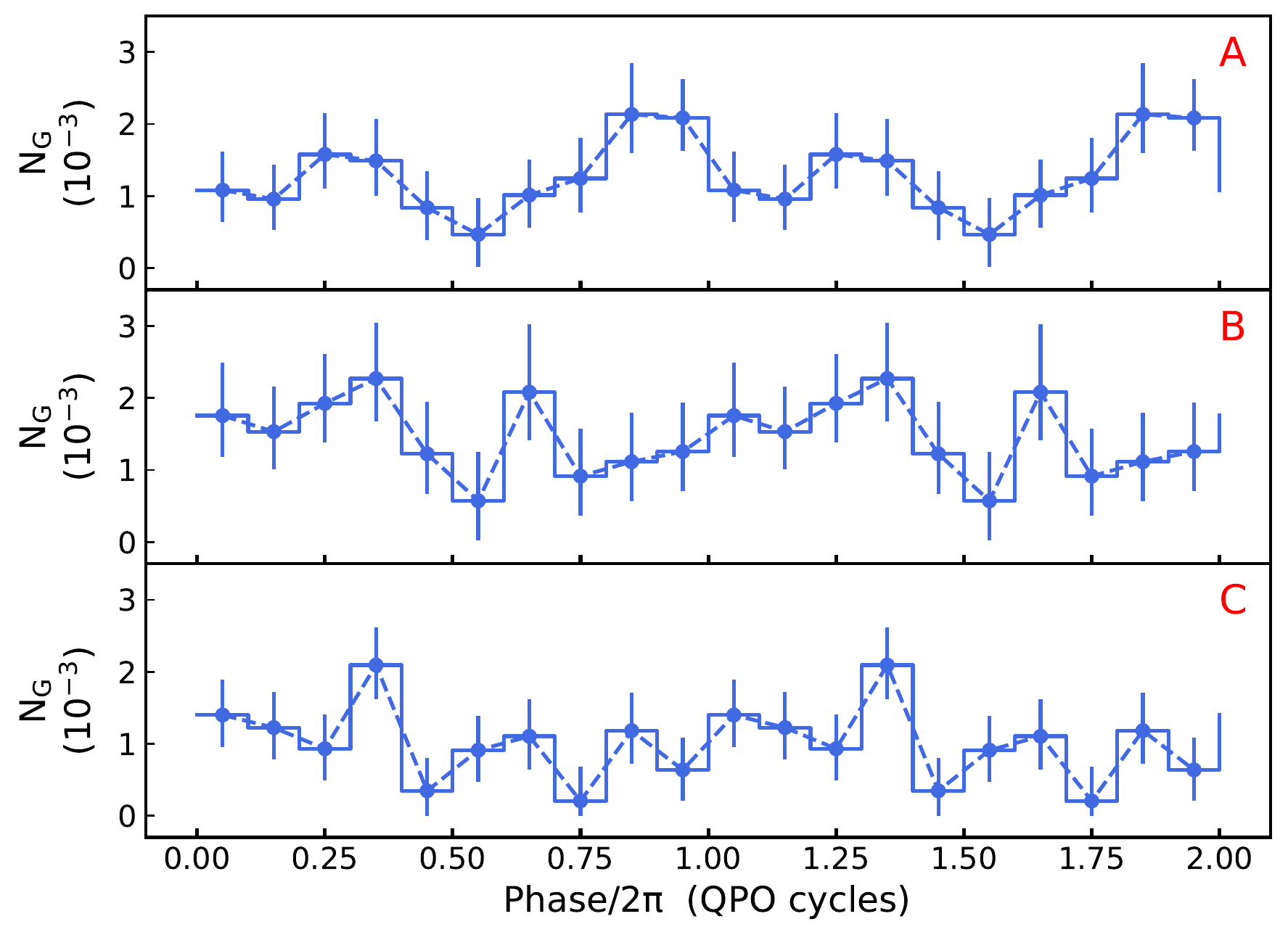}}
\caption{Left panel: Dynamic power spectrum of the second orbit of H1743--322. 
The ordinate is the period in seconds. 
The white dashed line indicates the period of the peak Lomb-Scargle power spectrum of the full dataset, which is the period we used to extract the phase-resolved spectra. 
The red horizontal line indicates the time interval of the beginning of the sliding window showing similar modulation of the Gaussian normalisation as found in the first orbit (panel A at the right panel). 
Right panel: Fitting result of the Gaussian normalisation for three selected sub-datasets. 
The start time of the sliding window are marked in the left panel. 
}
\label{p5}
\end{center}
\end{figure*}

We created phase-resolved spectra for each sub-datasets within a sliding data window of $\sim$1200 seconds, each time fitting the spectra with the same model, focusing on the fitting result of the Gaussian normalisation only. 
The width of the window ensured that we included 14 full QPO cycles. 
We found the same modulation of the Gaussian normalisation in all the sub-datasets of the first orbit; in the second orbit the modulation is present at the beginning of the observation (Figure~\ref{p5}A) which shows a variability of 1.7$\sigma$ significant, except that the phase of the modulation was shifted with respect to the phases of the first orbit, and appears to be absent in the rest of the observations (Figure~\ref{p5}). 
We show the DPS for the second orbit in the left panel of Figure~\ref{p5}, and in the right panels we show the results of the Gaussian normalisation for the three selected sub-datasets (marked with red triangles). 
Sub-panel A shows a modulation similar to what we have found in the first orbit, with the two peaks appearing at phase 0.3 and 0.9 of the QPO cycle. 
This modulation corresponds to the time interval marked with the red line in the left panel of the plot. 
Point B indicates the time of the second orbit at which the QPO period shows the largest deviation from the mean value of the full dataset. 
At point C, the period of the QPO returns back to the mean value. 

%%%%%%%%%%%%%%%%%%%%%%%%%%%%%%%%%%%%%%%%%%%%%%%%%%%%%%%%%%%%%%%%%%%%%%%%%%%%%
%%%%      Discussion                                                     %%%%
%%%%%%%%%%%%%%%%%%%%%%%%%%%%%%%%%%%%%%%%%%%%%%%%%%%%%%%%%%%%%%%%%%%%%%%%%%%%%
\section{Discussion}
\label{discussion}

The black-hole candidate H1743--322 showed a unique QPO at $\sim11$ mHz during its 2010 and 2011 outbursts, so far only detected in two \emph{RXTE} and one \emph{Chandra} observations of this source. 
Here we analysed the two \emph{RXTE} observations that show this mHz QPO. 
The frequency of the mHz QPO is the same in these two outbursts, whereas the frequency of a type-C QPO in these observations varies by a factor of two. 
We carried out phase-resolved spectroscopy on the period of the mHz QPO using the \emph{RXTE} observations, and found a modulation of the Gaussian line normalisation in the first orbit of the \emph{RXTE} data, with the line normalisations showing two maxima and two minima in one QPO cycle. 
This modulation is not affected by the continuum model. 
Similar, yet weaker modulation, is present at the beginning of the data of the second orbit, but this modulation disappeared as the observation progressed. 
The line flux (proportional to the Gaussian normalisation, since the line energy and width are linked across phases) modulation is very similar to what \cite{ingram2016} found for the type-C QPO. 
% We discussed the possible interpretation of the line flux modulation of this mHz QPO and the possible connection with a similar modulation of the line parameters with the period of the type-C QPOs in the following sections. 
We interpret the line flux modulation of this mHz QPO in terms of Lense-Thirring precession of the inner flow and discuss the possible connection with a similar modulation of the line parameters with the period of the type-C QPO in the following sections.

\subsection{Possible interpretation of the modulation of the line flux}

Our data show a modulation of the line flux with the 11 mHz QPO in the first orbit, 
reminiscent to the modulation of the Gaussian line centroid energy and flux with twice the frequency of the type-C QPO in this source \citep{ingram2016,ingram2017}. 
In that case the frequency of the type-C QPO was $\sim0.20$ to $\sim0.25$ Hz, about 20 times higher than the frequency of the mHz QPO we analysed here. 
\cite{ingram2016} combined five \emph{XMM-Newton} observations and found a modulation of the iron line centroid energy with a significance of 3.7$\sigma$, whereas the modulation of the line flux was only 1.3$\sigma$ significant in the other observations. 
For another observation, which they called anomalous (see \citealt{ingram2016}), the line flux shows a large amplitude modulation in phase with the modulation of the line energy, 
but anticorrelated with the modulation of the line energy seen in the other data sets. 
\cite{ingram2016} explained the red and blue-shift of the iron line as due to Lense-Thirring precession of the hot inner flow around the BH. 
The disc is truncated at some radius larger than the innermost stable circular orbit (ISCO), forming a large scale-height hot inner flow in this region, which is misaligned with respect to the accretion disc \citep{done2007}. 
Different azimuths of the disc will be illuminated by both the front and back part of the precessing inner flow, resulting in oscillations of the line energy with two maxima and two minima in one QPO cycle when the approaching and the receding sides of the disc are illuminated; 
the maximum of the line flux occurs when the illuminated part of the disc faces towards us. 

Our data show a modulation of the line flux that changes by a factor of $\sim4$ in the first orbit, which is similar to the case of the anomalous data set in \cite{ingram2016} where it was interpreted as due to the change of the ionisation state of the disc. 
According to \cite{ingram2016}, as the misalignment between the disc and inner flow caused by the precession motion of the inner flow changes, the disc ionisation state will also change within the QPO cycle, due to changes of the number of hard continuum photons that are intercepted by the disc; this in turn leads to a modulation of the line flux in phase with the modulation of the line energy. 
Unfortunately, the limited \emph{RXTE} energy resolution makes it impossible to resolve a possible energy shift of the emission line, so it is unclear whether or how the line energy changes with the line flux in our observation.

A weaker modulation likely present at the beginning of the second {\emph RXTE} orbit (Figure~\ref{p5}A), that disappeared as the observation progressed further, suggests that the process that drives the modulation is unstable. 
This could be due to an unstable inner flow geometry, e.g., if the misalignment angle between the disc and the inner flow decreases, the disc ionisation would also decrease, leading to a lack of line flux modulation. 
We also noticed that, in the data of the second orbit, the line flux modulation has a phase shift compared to the modulation in the first orbit. 
Fits to these two modulations with a sine function yields a phase difference of 0.2 of the oscillation period, suggesting a possible change of the inner flow geometry between these two orbits. 
For a 10$M_{\odot}$ BH, the structure, for instance two hot spots in the inner flow, that produce this $\sim11$ mHz QPO would move $\sim1000$ km within $\sim2400$ seconds interval between the first two orbits of {\emph RXTE} observation.

\subsection{Line flux modulation for different types of QPOs}

\begin{table*}
\caption{mHz QPOs identified from Black-Hole candidates in previous studies.}
\begin{tabular}{l c c c c c}
\hline
Source & QPO frequency & rms(\%) & X-ray state$^{a}$ & inclination & note\\
\hline
H1743--322 & 11 mHz & $\sim3$ & (rising) LHS & $\sim60^{\circ}-70^{\circ}$ & 1 \\
% H1743-322 & 11 mHz \& 0.919 Hz & $\sim3$ \& 12 & (rising) LHS & $\sim60^{\circ}-70^{\circ}$ & 1 \\
 % & 11 mHz \& 0.424 Hz & $\sim3$ \& 11 & (rising) LHS &  & \\
V404 Cyg & 18 mHz & $18\pm2$(\emph{Swift}) & (rising) LHS & $67^{\circ}$$_{-1}^{+3}$ & 2 \\
 & 73 mHz & $27\pm3$ & (rising) LHS &  &  \\
% V404 Cyg & 18 mHz & $18\pm2$(\emph{Swift}) & (rising) LHS & $67^{\circ}$$_{-1}^{+3}$ & 2 \\
 % & 73 mHz \& 1.03 Hz & $27\pm3$ \& $46\pm2$ & (rising) LHS &  &  \\
 & 136 mHz & $8\pm2$ & SIMS &  &  \\
IC 10 X-1 & 6.3 mHz & $11\pm3$ & - & $\ge63^{\circ}$ & 3 \\
Swift J1357.2--0933 & 5.9 mHz & $12\pm3$ & LHS & $\gtrsim70^{\circ}$ & 4 \\
% \hline
XTE J1118+480 & 69-159 mHz & $\sim4-10$ & LHS & $68^{\circ}-79^{\circ}$ & 5 \\
% \hline
LMC X-1 & $\sim$27 mHz & $\sim1-2$ & HSS & $36.38^{\circ}\pm1.92^{\circ}$ & 6 \\
% \hline
Cygnus X-3 & 8.5 mHz \& 30 mHz & $\lesssim3$ & HSS & likely low inclination & 7 \\
 & 9 mHz, 21 mHz \& 31 mHz & $\lesssim2$ & HSS &  &  \\
% \hline
MAXI J1820+070 & 30-42 mHz & - & LHS/HIMS & - & 8 \\
\hline
\end{tabular}
\begin{tablenotes}
% \small
\item $^{a}$LHS refers to Low-Hard state, HSS refers to High-Soft state, SIMS refers to Soft-Intermediate state, HIMS refers to High-Intermediate state. 
\item Notes: 1. Type-C QPOs have been detected simultaneously with the $\sim11$ mHz QPOs in the same observations during the 2010 and 2011 outbursts at 0.919 Hz (rms 12\%) and 0.424 Hz (rms 11\%), respectively \citep{diego2012}; 2. A LFQPO at 1.03 Hz (rms 46\%) has been detected simultaneously with the 73 mHz QPO in the {\emph Chandra} observation. The fractional rms amplitude is strongly energy-dependent, there is strong radio activity coincident with the X-ray flaring \citep{huppenkothen2017,mooley2015}; 3. The fractional rms amplitude of the QPO is roughly energy-independent in the $0.3-1.5$ keV energy range \citep{pasham2013,laycock2015}; 4. The source shows optical dips at frequencies similar to the frequency of the mHz QPOs \citep{padilla2014}; 5. Similar QPOs feature were observed in X-rays, EUV and optical \citep{wood2000,khargharia2013,revnivtsev2000,haswell2000}; 6. The presence of the mHz QPOs and the broad iron line in the X-ray emission appears to depend on the presence of a strong power-law component \citep{alam2014,orosz2009}; 7. The presence of mHz QPOs is associated to major radio flaring events \citep{koljonen2011,vilhu2013}; 8. \cite{mereminskiy2018}. 
\end{tablenotes}
\label{t2}
\end{table*}

\cite{ingram2016,ingram2017} found that the centroid energy and flux of the iron line oscillate at twice the frequency of the type-C QPO, which at that time was $\sim$0.20--0.25 Hz; here we found that the line flux is modulated at twice the frequency of the QPO at $\sim$11 mHz. 
It is then quite interesting to know how the iron line flux can be modulated by different types of QPOs. 
Type-C QPOs were also detected in our observations; \cite{diego2012} reported the type-C QPO and its second harmonic at 0.9 Hz and 1.8 Hz in the 2010 observation, and at 0.4 Hz and 0.8 Hz in the 2011 observation (see also Figure~\ref{p1}). 
% Since the frequency of the type-C QPO in these two observations differs by a factor of $\sim$2, even though the mHz QPO shows a similar frequency in both observations, 
The mHz QPO shows a similar frequency in both observations, however, the frequency of the type-C QPO in these two observations differs by a factor of $\sim$2; 
because of this \cite{diego2012} concluded that the mHz and type-C QPO may be produced by different mechanisms. 

Assuming that the modulation of the iron line energy/normalization found by \cite{ingram2016} is always present (including at the time of the observations analyzed in this paper), our results imply that the line flux modulation remains the same even if the QPO frequency changes by more than an order of magnitude. 
Under the interpretation of Lense-Thirring precession \citep{ingram2016}, our results could be explained if the mHz QPO is due to a precessing torus with a radius that is $\sim3-5$ times larger than the radius of the precessing inner flow predicted by the frequency of the type-C QPOs.

The possibility of the existence of this two-components inner flow is interesting, yet it still needs to be tested. 
\cite{qu2010} and \cite{yan2012} found that the centroid frequency of the LF QPOs in GRS~1915+105 decreases with energy when the QPO frequency is below $\sim2$ Hz, but increases with energy when the QPO is above $\sim2$ Hz. 
A similar behaviour has also been reported for XTE J1550--564 \citep{li2013a} and H1743--322 \citep{li2013b}: in both sources the frequency increases with energy when the QPO frequency is higher than a certain value (this turn-over frequency is different for different sources). 
However, GRS~1915+105 is the only source that shows decreases of QPO frequency with energy, which is still unexplained by current models. 
To explain these results, \cite{eijnden2016a} proposed a geometric toy model of differential precession \citep{ingram2009} by considering the inner flow as two separate components: an outer and an inner half, with the latter always producing a QPO at higher frequency than the former. 
For higher full-band QPO frequencies, the spectrum of the inner half is harder than that of the outer half, whereas the outer half is harder when the full-band QPO frequencies are lower than the turn-over frequency. 
This spectral evolution could either be due to the ionisation state changes of the reflected component which dominated the outer part, or to the presence of an extra cooling process very close to the black hole.

This is consistent with what we have observed here, as the phase-resolved spectroscopy of H1743--322 shows that the variability of the spectra among different phases are dominated by the soft photons for this mHz QPO, which are likely coming from the outer half of the inner flow. 
\cite{ingram2016} showed that in H1743--322 the power-law index changes from $\sim1.26$ to $\sim1.29$ during the $\sim0.2$ Hz QPO cycles, which is significantly harder than the values found in the average spectra of our observations for the mHz QPO, $1.60\pm0.01$ for orbit 1, $1.59\pm0.01$ for orbit 2, and $1.53\pm0.01$ for orbit 3. 
These results indicate that the $\sim11$ mHz QPOs may originate in the outer (cooler) region of the precessing inner flow, while the type-C QPOs originate in the hotter inner half of that flow.

The detailed structure of an inner flow that can produce these two independent LF QPOs simultaneously needs to be investigated. 
% The Lense-Thirring frequency is given by \cite{stella1998} as 
% ${\nu_{\rm LT}^{}} = 8\pi^2 I \nu_{\rm K}^2\nu_{\rm s}/(M c^2)$, 
% where $M$ is the mass of BH, $I$ is the moment of inertia, $\nu_{\rm K}$ and $\nu_{\rm s}$ are the Keplerian frequency and spin frequency, respectively.  
For the case of GRS~1915+105, \cite{qu2010} found that the largest variation of the QPO frequency with energy occurs when the QPO is at $\sim6$ Hz, where it changes from $\sim5.9$ Hz in the 1.94--5.12 keV band to $\sim6.7$ Hz in the 18.09--38.44 keV band. 
If $R_{\rm in}$ is the distance from the BH to the inner half precessing inner flow, for GRS~1915+105 the outer half could extend up to a radius of 1.04$R_{\rm in}$. 
For the case of H1743--322, the mass of the BH is thought to be $\sim$10 M$_{\odot}$ with a spin parameter of 0.2 \citep{steiner2012}. 
If the type-C QPO is due to Lense-Thirring precession, the inner half of the inner flow would have a radius of 10 $r_{\rm g}$ during the 2010 outburst, and 13 $r_{\rm g}$ during the 2011 outburst;  
the frequency of the mHz QPO is stable in these two outbursts, and the outer half of the inner flow would have a radius of 46 $r_{\rm g}$. 
Compared to GRS~1915+105, this requires a large truncation radius to form such an extended structure. 
\cite{diego2012} found that the frequency of this mHz QPO changes by less than $\sim$1.5 mHz during the 60 ks \emph{Chandra} observation in the 2010 outburst, 
whereas \cite{ingram2016} found that the type-C QPO frequency changed from $\sim0.2$ Hz to $\sim0.25$ Hz during the \emph{XMM-Newton} observation in the 2014 outburst. 
Those facts indicate that, if the mHz QPO and the type-C QPO may both originate from the precessing inner flow (at different radii), the structure of these two components should be relatively independent, and the extreme environment of the inner region near the BH must be much more unstable and complicated than the outer region. 

Besides the $\sim11$ mHz QPOs in H1743--322 \citep{diego2012}, several other sources have shown QPOs with frequency in the mHz range (see Table~\ref{t2}). 
Unlike the typical LFQPO in BHXRBs, these mHz QPOs appear at frequencies that are lower than most of the type-A ($\sim$ 8 Hz), type-B ($\sim$5-6 Hz) and type-C ($\sim$0.1-15 Hz) QPOs. 
Except for LMC X-1 and Cygnus X-3, the mHz QPOs in these systems have relatively large fractional rms amplitude and appear during the Low-Hard State (LHS), which distinguishes them from the so-called `heartbeat' QPOs found in GRS~1915+105 \citep{belloni2000} and IGR J17091--3624 \citep{altamirano2011} in the mHz range. 

Most of the sources listed in Table~\ref{t2} are believed to have high orbital inclinations, which appears to connect these mHz QPOs to the so-called 1-Hz QPO observed in dipping neutron star X-ray binaries. 
The frequency of the 1-Hz QPO has been observed to vary between 0.4 and 3.0 Hz \citep{homan1999,homan2012}; 
these QPOs have been explained as obscuration by a radiation-driven warping of the accretion disc \citep{jonker1999,pringle1996}, or to a precessing inner accretion disc \citep{homan2012}. 
% the obscuring effect of the disk instabilities in the boundary layers \citep{jonker1999,bhattacharyya2006,homan2012}. 
From the list of sources in Table~\ref{t2}, Swift J1357.2--0933 and XTE J1118+480 have shown both optical dips at a similar frequency as the mHz QPOs \citep{padilla2014,wood2000}; this raises the possibility that the structure obscuring the optical emission may also cause the mHz QPO. 
However, in Swift J1357.2--0933 the optical dips were detected in several occasions at a frequency that decreased as the X-ray luminosity decreased, whereas the X-ray mHz QPOs were only detected in one observation and were significantly absent in the other cases in which optical dips were observed \citep{padilla2014}. 
Therefore, there is no direct evidence that these two phenomena are physical related. 
Given the presence of the mHz QPOs associated with radio flaring events in V404 Cyg and Cygnus X-3 \citep{huppenkothen2017,koljonen2011}, an alternative explanation is that these QPOs are produced at the base of a precessing jet \citep{markoff2005,kalamkar2016}. 

LMC X-1 and Cygnus X-3 share more common features, but appear to be part of a different group compared with the other six sources with mHz QPOs listed in Table~\ref{t2}: in these two sources the mHz QPOs have relative low fractional rms amplitudes ($\lesssim$ 3\%), the QPOs appear in the High-Soft State (HSS), and the two systems are believed to have low orbital inclinations. 
Those facts suggest that the mHz QPOs in low-inclination systems would either have a different origin compared with mHz QPOs in high-inclination systems or, just like the type-C QPOs, the mHz QPOs may also have a geometric origin which leads to the inclination dependent properties: the oscillations are stronger for high-inclination than for low-inclination sources \citep{motta2015,eijnden2017}. 
If this is true, then it should not be a surprise that the line flux modulation with the mHz QPO frequency is similar to that of the type-C QPOs. 
However, H1743--322 is exceptional in this respect, as the strong hard dips found in the light curves in this source suggest a high system inclination ($\sim$60$^{\circ}$-70$^{\circ}$, \citealt{homan2005}) but with a relative low rms amplitude ($\sim$ 3\%, \citealt{diego2012}). 
\cite{steiner2012} measured the jets inclination of H1743--322 as $75^{\circ}\pm3^{\circ}$, indicating that there is possibly a very large misalignment between the BH equatorial plane and the accretion disc compared with other sources, and this may be the reason that makes the geometry of the inner flow in H1743--322 different from that in the other high inclination systems. 

We also noticed that in the observations of H1743--322 and V404 Cyg that show two distinct LFQPOs simultaneously at, respectively, 10--70 mHz and 0.5--1 Hz, the QPOs at 0.5--1 Hz have a significantly higher rms amplitude than the QPOs at mHz frequencies (see Table~\ref{t2} Notes 1 and 2; there is not enough information about the LFQPOs in Cygnus X-3 yet to draw a similar conclusion). 
It remains to be seen whether this trend holds in all cases. 

Since the list in Table~\ref{t2} only includes eight sources with mHz QPO detection, while there is not much information yet available for the newly discovered BHC MAXI J1820+070, the reliability of the geometric origin of these mHz QPOs is still unclear. 
As the other interpretations still have difficulties in explaining the line flux modulation of the mHz QPOs, the Lense-Thirring precession of the inner flow at large radii is more likely to be the cause of the mHz QPOs in H1743--322. 
More observations of these LFQPOs are required to reveal the possible relations between the different types QPOs, to help us better understanding their physical origins and, perhaps, the geometry of the accreting flow close to the BH.

%%%%%%%%%%%%%%%%%%%%%%%%%%%%%%%%%%%%%%%%%%%%%%%%%%%%%%%
%%
%%
%% UP TO HERE
%%
%%
%%%%%%%%%%%%%%%%%%%%%%%%%%%%%%%%%%%%%%%%%%%%%%%%%%%%%%%

\section*{Acknowledgments}

This research has made use of data and software provided by the High Energy Astrophysics Science Archive Research Center (HEASARC), which is a service of the Astrophysics Science Division at NASA/GSFC and the High Energy Astrophysics Division of the Smithsonian Astrophysical Observatory. 
ZC thanks Jamie M.C. Court for the helpful discussion about the algorithm in creating GTIs of each phase. 
DA acknowledges support from the Royal Society. 
A.B acknowledges support from the Royal Society and SERB (Science \& Engineering Research Board, India) through Newton-Bhabha Fund.

%%%%%%%%%%%%%%%%%%%%%%%%%%%%%%%%%%%%%%%%%%%%%%%%%%%%%%%%%%%%%%%%%%%%%%%%%%%%%
%%%%    References                                                       %%%%
%%%%%%%%%%%%%%%%%%%%%%%%%%%%%%%%%%%%%%%%%%%%%%%%%%%%%%%%%%%%%%%%%%%%%%%%%%%%%
\bibliographystyle{mn2e}
\bibliography{paper_7.bbl}

%%%%%%%%%%%%%%%%%%%%%%%%%%%%%%%%%%%%%%%%%%%%%%%%%%%%%%%%%%%%%%%%%%%%%%%%%%%%%
%%%%    Appendix                                                         %%%%
%%%%%%%%%%%%%%%%%%%%%%%%%%%%%%%%%%%%%%%%%%%%%%%%%%%%%%%%%%%%%%%%%%%%%%%%%%%%%
% \begin{appendix}
% \end{appendix}

%%%%%%%%%%%%%%%%%%%%%%%%%%%%%%%%%%%%%%%%%%%%%%%%%%%%%%%%%%%%%%%%%%%%%%%%%%%%%
%%%%%%%%%%%%%%%%%%%%%%%%%%%%%%%%%%%%%%%%%%%%%%%%%%%%%%%%%%%%%%%%%%%%%%%%%%%%%
\vspace{1cm}
\footnotesize{This paper was typeset using a \LaTeX\ file prepared by the 
author}

%%%%%%%%%%%%%%%%%%%%%%%%%%%%%%%%%%%%%%%%%%%%%%%%%%%%%%%%%%%%%%%%%%%%%%%%%%%%%
%%%%%%%%%%%%%%%%%%%%%%%%%%%%%%%%%%%%%%%%%%%%%%%%%%%%%%%%%%%%%%%%%%%%%%%%%%%%%

\end{document}